\documentclass[manuscript]{aastex}

\usepackage{tikz}
\usepackage{amsmath}

\usepackage{graphicx}

\usepackage{pdflscape}

\usepackage{hyperref}

\slugcomment{\textbf{The Astronomical Journal} }

\shorttitle{2017 S5}
\shortauthors{Jewitt}

\begin{document}

\title{Active Asteroid P/2017 S5 (ATLAS)}

%
\author{David Jewitt$^{1,2}$,
Yoonyoung Kim$^3$,  
Jayadev Rajagopal$^4$, Susan Ridgway$^4$, Ralf Kotulla$^5$,  Wilson Liu$^4$, 
Max Mutchler$^6$, Jing Li$^{1}$,  Harold Weaver$^7$, \\
and
Stephen Larson$^8$ 
}
\affil{$^1$Department of Earth, Planetary and Space Sciences, UCLA, 
595 Charles Young Drive East,  Los Angeles, CA 90095-1567\\
$^2$Dept.~of Physics and Astronomy, University of California at Los Angeles, \\
430 Portola Plaza, Box 951547, Los Angeles, CA 90095-1547\\
$^3$ Max Planck Institute for Solar System Research, Justus-von-Liebig-Weg 3, 37077 G\"ottingen, Germany\\
$^4$ NOAO, 950 North Cherry Ave., Tucson, AZ 85719 \\
$^5$ Department of Astronomy, University of Wisconsin-Madison, 475 N. Charter St., Madison, WI 53706 \\
$^6$ Space Telescope Science Institute, 3700 San Martin Drive, Baltimore, MD 21218 \\
$^7$ The Johns Hopkins University Applied Physics Laboratory, 11100 Johns Hopkins Road, Laurel, MD 20723  \\
$^8$ Lunar and Planetary Laboratory, University of Arizona, 1629 E. University Boulevard, Tucson, AZ 85721-0092 \\
}

\email{jewitt@ucla.edu}

\begin{abstract}
Observations of active asteroid P/2017 S5 when near perihelion reveal  the ejection of large (10$^2$ to 10$^4$ $\mu$m) particles  at 0.2 to 2 m s$^{-1}$ speeds, with estimated mass-loss rates of a few kg s$^{-1}$.  The protracted nature of the mass loss (continuous over $\sim$150 days) is compatible with a sublimation origin, meaning that this object is likely an ice-bearing main-belt comet.  Equilibrium sublimation of exposed water ice covering as little as 0.1 km$^2$ can match the data. Observations a year after perihelion show the object in an inactive state from which we deduce a nucleus effective radius 450$_{-60}^{+100}$ m (albedo 0.06$\pm$0.02 assumed).  The gravitational escape speed from a body of this size is just $\sim$0.3 m s$^{-1}$, comparable to the inferred ejection speed of the dust.  Time-series photometry provides tentative evidence for rapid rotation (lightcurve period 1.4 hour) that may also play a role in the loss of mass and which, if real, is a likely consequence of spin-up by sublimation torques.  P/2017 S5 shares both physical and orbital similarities with the split active asteroid pair P/2016 J1-A and J1-B, and all three objects are likely members of the $\sim$7 Myr old, collisionally produced, Theobalda family.

\end{abstract}


\section{INTRODUCTION}
\label{intro}

Small solar system body P/2017 S5 (hereafter ``S5'') was discovered by Heinze et al.~(2017) on UT 2017 September 27 using the  0.5 m ATLAS survey telescope on Haleakala.  The orbit  has semimajor axis $a$ = 3.171 AU, eccentricity $e$ = 0.313, inclination $i$ = 11.8\degr, giving an asteroid-like Tisserand parameter with respect to Jupiter, $T_J$ = 3.09.   These orbital properties, combined with a comet-like appearance, qualify S5 as a member of the active asteroids population (Jewitt et al.~2015).  Perihelion occurred shortly before discovery on 2017-Jul-28 (DOY = 210) at distance $q$ = 2.178 AU.

The central question about the active asteroids as a group is ``what drives the observed mass loss?''.  The range of processes  identified to date includes impact, breakup, rotational mass-shedding and the sublimation of near-surface ice.  Objects in the latter category are labelled ``main-belt comets'' (Hsieh and Jewitt 2006), and form a distinct sub-set of the active asteroids defined by the recurrence of their activity. In the case of S5, Novakovic (2018) reported a likely dynamical association with the Theobalda asteroid family.  With mean semimajor axis $a \sim$ 3.175 AU, this family has a model dynamical  age $\tau_d$ = 7$\pm$2 Myr and is thought to have resulted from the impact-caused disruption of a  78$\pm$9 km diameter parent body (Novakovic 2010, Hsieh et al.~2018).  Asteroid 778 Theobalda itself has a diameter of $\sim$56 km and a geometric albedo of $\sim$0.08 (Mainzer et al.~2016).  The outer belt location and recent origin of the family suggest that S5 could be an ice-bearing fragment from the collision, perhaps recently activated by a separate collision with a boulder-sized projectile, although distinct evidence for this possibility remains to be found.

In this paper, we combine new high-resolution observations from the Hubble Space Telescope with sensitive ground-based images from the Wisconsin Indiana Yale NOAO ``WIYN'' telescope  in order  to characterize the nucleus and the activity of S5.

\section{OBSERVATIONS}
Observations were  taken on UT 2017 November 27 at the 3.5 meter diameter WIYN telescope, located at Kitt Peak National Observatory in Arizona.  We used the One Degree Imager (ODI) camera mounted at the Nasmyth focus  with a native image scale of 0.11\arcsec~pixel$^{-1}$, the latter rebinned to 0.25\arcsec~pixel$^{-1}$ in the images presented here (Harbeck et al, 2014).  Observations were taken through the Sloan  r' (central wavelength $\lambda_c$ = 6250\AA, full-width at half-maximum (FWHM) = 1400\AA) filter with 180 s integration time.  Seeing was variable in the range 0.8\arcsec~to 1.2\arcsec. Data reduction used the ``Quickreduce'' pipeline (Kotulla 2014), and photometric calibration of the data was made with reference to the Sloan DR 14 database (Blanton et al.~2017).

We used the 2.4 m Hubble Space Telescope (HST)  to image S5 on UT 2018 September 08.  The WFC3 camera yields an image scale 0.04\arcsec~pixel$^{-1}$ over a field of view 162\arcsec~square.  We used the F350LP filter (central wavelength $\lambda_C =$ 6230\AA~when used on a source with a sun-like spectrum and  FWHM = 4758\AA). In each of two consecutive orbits we obtained a sequence of five integrations each of 360 s duration, for a total on-source time of 3600 s.  One of the images was severely compromised by a cosmic ray strike and we reject it from further consideration.  Four additional integrations of 10 s each were obtained as a precaution against potential saturation of the nucleus.  Since the nucleus was not saturated, we have not used these short integrations in the present investigation.   
To calibrate the photometry we assumed that, as indicated by the HST on-line Exposure Time Calculator tool (\url{http://etc.stsci.edu/}),  a $V$ = 0 magnitude solar-type source would give a count rate 4.71$\times$10$^{10}$ s$^{-1}$.   

In addition to the WIYN and HST observations, we identified an archival observation from the DECam survey taken on UT 2017 October 25.  The image is a single 103 s integration  through a Sloan z filter ($\lambda_c =$ 9097\AA, FWHM = 1370\AA) with an image scale 0.27\arcsec~pixel$^{-1}$ and seeing 0.9\arcsec~FWHM.  The object is readily identifiable by its appearance and by the lack of any counterpart at the same location in archival data taken at other times.  

\begin{figure}
\epsscale{0.99}
\plotone{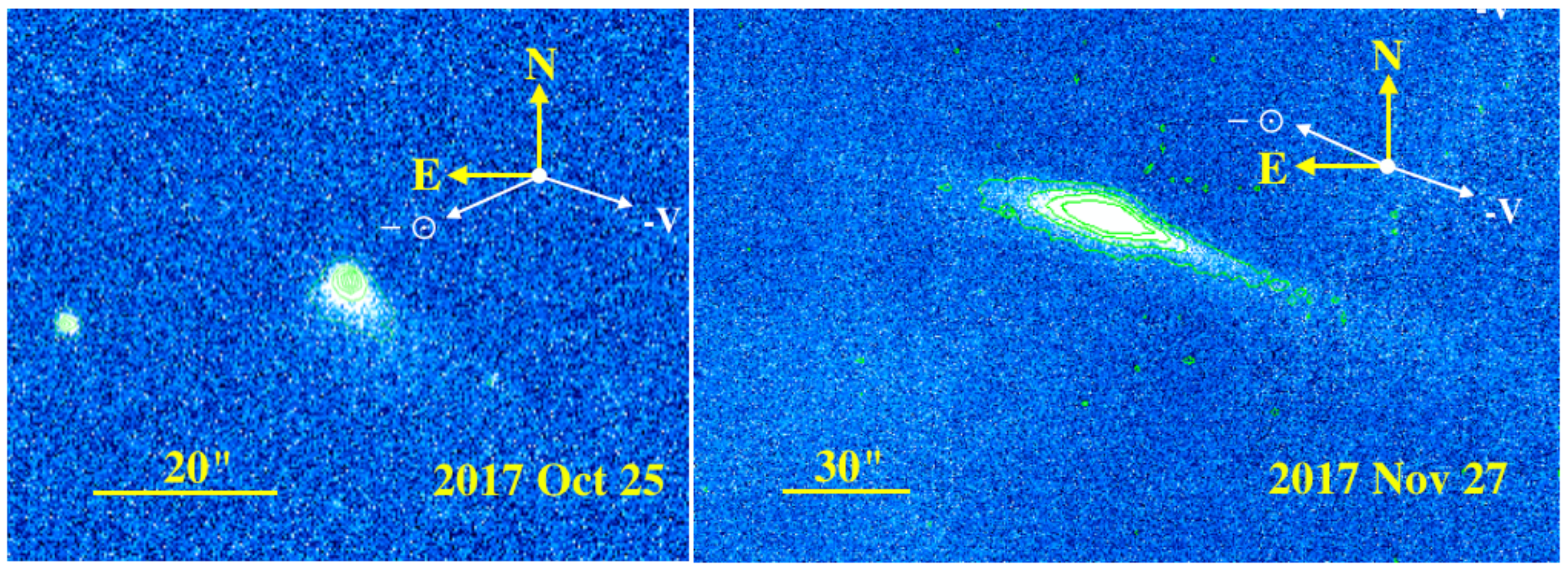}
\caption{(left): CTIO 4 m image taken UT 2017 October 25. (right): WIYN 3.5 m image taken UT 2017 November 27.  The cardinal directions,  the position angles of the anti-solar vector (marked $-\odot$) and the negative projected heliocentric velocity vector ($-V$), and a scale bar, are shown for each image.  
\label{image1}}
\end{figure}

Magnitudes from WIYN and DECam were transformed from  the Sloan system to Johnson-Cousins V using the relations given by Jordi et al.~(2006) assuming solar colors (V-R = 0.35, R-I = 0.33) for S5.     Composite images from these telescopes are shown in  Figure (\ref{image1}).

\section{DISCUSSION}

\subsection{Nucleus}

\begin{figure}
\epsscale{0.99}
\plotone{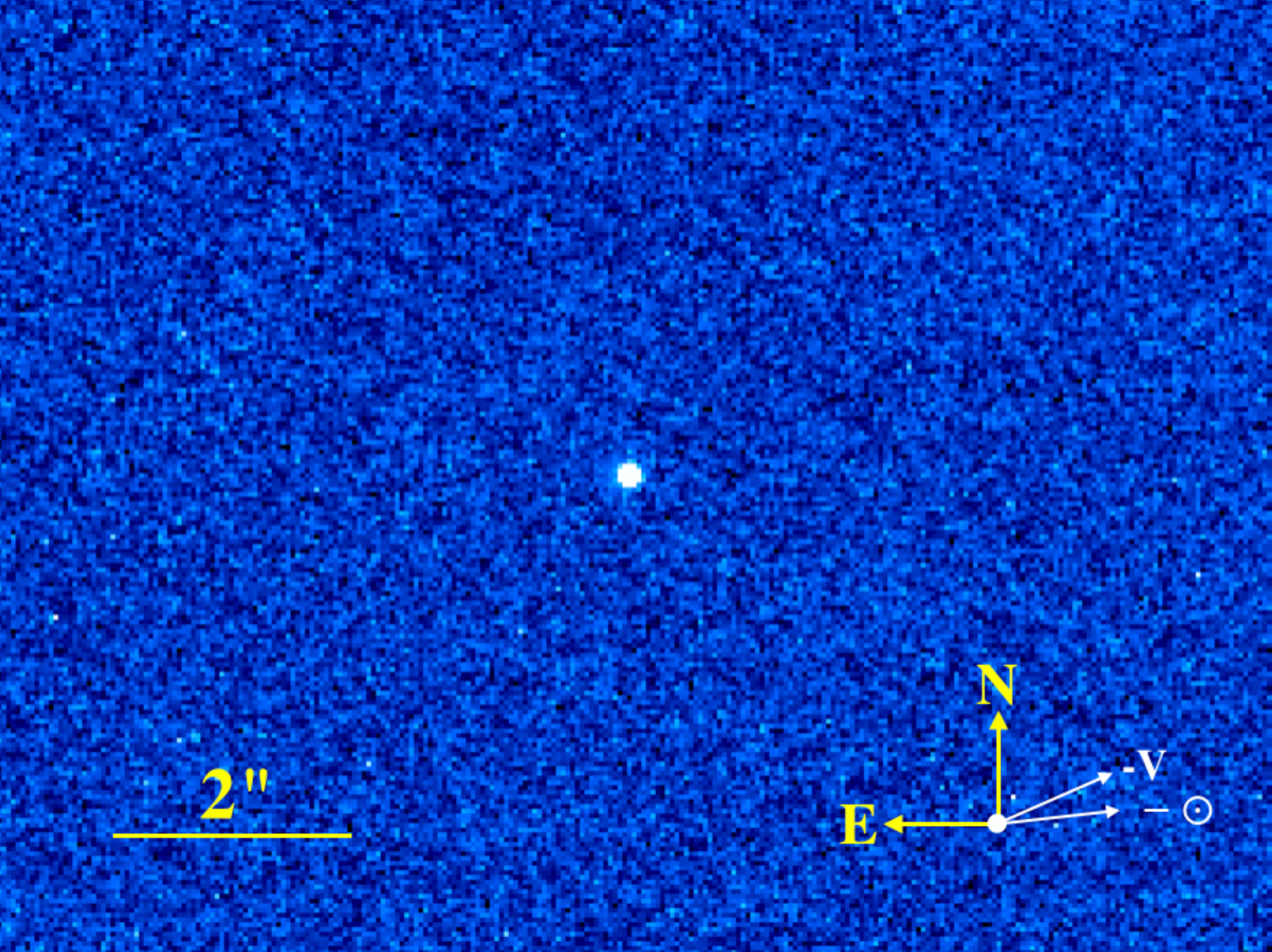}
\caption{Composite HST 3240 s image from UT 2018 September 07.  The cardinal directions, and the position angles of the anti-solar vector (marked $-\odot$) and the negative projected heliocentric velocity vector ($-V$) are shown, along with a scale bar.
\label{image2}}
\end{figure}

The HST images were registered and combined into a single deep image, shown in Figure (\ref{image2}).  The FWHM of this image is $\theta_{FWHM} = 0.09\pm0.01$\arcsec,  consistent with the 0.08\arcsec~two-pixel Nyquist image width delivered by WFC3.  Accordingly, we proceed on the assumption that the measured brightness of S5 in the HST data is a measure of the albedo and cross-section of the nucleus.

The apparent magnitude determined within a 0.2\arcsec~radius projected circle, with sky subtraction from a contiguous annulus extending to 4.4\arcsec~outer radius, is $V$ = 25.05$\pm$0.03 (standard error on the mean of nine measurements).  With an assumed  phase function parameter (i.e.~$G$ = 0.15 in the photometric system of Bowell et al.~1989), the correction to unit heliocentric and geocentric distances and to 0\degr~phase angle is -6.14 magnitudes, giving absolute magnitude $H$ = 18.91.  Then, the effective scattering cross-section, $C_e$, is given by

\begin{equation}
p_V C_e = 2.25\times 10^{22} \pi 10^{0.4(V_{\odot} - H)}
\label{invsq}
\end{equation}

\noindent where $p_V$ is the geometric albedo and $V_{\odot}$ = -26.74 is the apparent V magnitude of the Sun (Cayrel de Strobel~1996).   The geometric albedo of S5 is unmeasured.  In its place, we adopt $p_V = 0.06\pm0.02$, the mean albedo of the Theobalda family asteroids (Masiero et al.~2013).  Then, substituting $H$ = 18.91 we obtain $C_e$ = 6.5$\times 10^5$ m$^2$.  The radius of an equal-area circle is $r_e = (C_e/\pi)^{1/2}$, giving $r_e$ = 450 m.  The formal statistical error from the scatter in the photometry is $\pm$3\% in the cross-section and only 1.5\% in the radius, but the real uncertainty is  larger, non-random and difficult to specify.  As a minimum estimate,  we take albedos $\pm1\sigma$ from the Theobalda family mean, to find that the effective radius lies in the range $r_e = 450_{-60}^{+100}$ m.  Evidently, S5 is a very small body, comparable in size to the smallest of the active asteroids for which meaningful size estimates exist (Jewitt et al.~2015).  Assuming a spherical shape and ignoring rotation, the gravitational escape speed is $V_e = (8\pi G \rho/3)^{1/2} r_e$, where $G = 6.67\times 10^{-11}$ N kg$^{-2}$ m$^2$ is the gravitational constant and $\rho$ is the density of S5.  With nominal density $\rho$ = 1000 kg m$^{-3}$, we find $V_e$ = 0.3 m s$^{-1}$.  The density and escape speed  could be smaller still, if S5 retains a fragmented, internally porous and therefore less dense structure. The escape speed would also be smaller if assisted by nucleus rotation, especially near the tips of a prolate body in rotation about its minor axis.  Thus, we regard $V_e$ = 0.3 m s$^{-1}$ only as a crude estimate of the escape speed from S5, perhaps accurate to within a factor of  two.

\begin{figure}
\epsscale{0.99}
\plotone{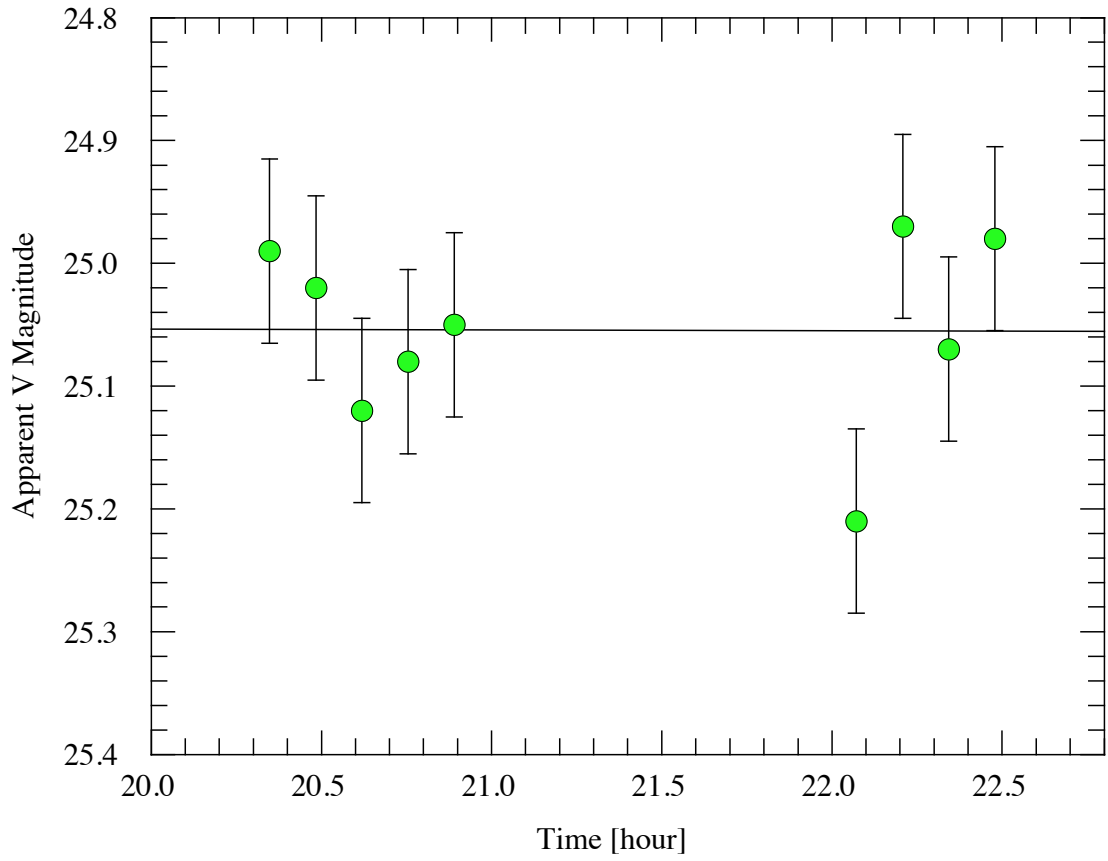}
\caption{Photometry from Table \ref{photometry1}.  The horizontal line is a least-squares fit to the data, showing the absence of a significant gradient.  
\label{lightcurve}}
\end{figure}

To search for rotational modulation of the scattered light, we measured each of the HST images individually within the 0.2\arcsec~radius aperture. The results are listed in Table (\ref{photometry1}) and plotted as a function of time in Figure (\ref{lightcurve}).  Within the (considerable) photometric uncertainties, we find no evidence for rotational variation of the scattered light over the $\sim$ 2 hour period of observations.  With no measurable lightcurve, there is no unique interpretation of the rotational state of the nucleus.  As judged from the HST data alone, S5 could be rotating very slowly (period $\gg$2 hr), or could be rotationally symmetric so that the lightcurve range is very small, or its rotation pole could be fortuitously close to the line of sight, so suppressing any rotational variation.  Conceivably, all of these things could be true.  All we can conclude from  Figure (\ref{lightcurve}) is that the data provide no evidence for rapid rotation in S5 and so cannot be used to argue that the observed activity is related in any way to the rotational state.  Later (in Section \ref{dust}), we describe photometric variations of S5 when in the active state which might be related to nucleus rotation, albeit indirectly.

 Slowly-escaping companions are thought to be a common product of asteroid rotational breakup (e.g.~Boldrin et al.~2016) leading, eventually, to the formation of asteroid pairs (Pravec et al.~2010).   Inspection of the composite HST image shows a number of very faint objects that superficially resemble the point spread function and which differ morphologically from cosmic ray strikes (for example, at the center top edge of the frame in Figure \ref{image2}).  We considered the possibility that one or more of these might be co-moving companions to the main nucleus of S5. To test this possibility, we combined the five 360 s exposures from the first orbit into a single image and blinked them against the four 360 s exposures from the second orbit combined into another image (one image from the second orbit was rejected because of a severe cosmic ray strike).  A real co-moving object should appear at the same location relative to the main nucleus in both combined images, but none of the candidates survived this simple test.  We conclude that the data provide no evidence for co-moving companions to the main body.  

To set a limit to the size of unseen secondary objects, we used the on-line Exposure Time Calculator to find that, in a 4$\times$360 s integration, signal-to-noise ratio SNR = 3 is reached at magnitude $V$ = 27.1.  This underestimates the limiting magnitude (by about 0.4 magnitudes) in the full nine-image combined frame, but provides a useful, conservative limit to the possible brightness of any companion object.  The corresponding limiting absolute magnitude is $H >$ 21.0 and, by Equation (\ref{invsq}), the upper limit to the radius is $r_e <$  170 m (geometric albedo $p_V$ = 0.06 assumed).

\subsection{Dust}
\label{dust}

We measured the  images of S5 in the active state from UT 2017 October and November using a circular aperture of fixed projected radius 25,000 km, so as to be sure that we are comparing the same volume around the nucleus on these dates.  Sky subtraction was determined from the median signal in concentric annuli with radii 100 pixels larger than the photometry apertures, in each case.  The resulting magnitudes are listed in Table (\ref{photometry2}), where it may be seen that the apparent magnitude faded by $\sim$0.4 magnitudes from 2017 October to November, but the absolute magnitude brightened by $\sim$0.4 magnitudes, showing the continued ejection of dust.  Both absolute magnitudes in the active state are brighter than the nucleus $H = 18.91$ by $\gtrsim$4 magnitudes, showing that  $\gtrsim$95\% of the cross-section in late 2017 was contributed by dust.  We also determined absolute magnitudes from the photometry by Borysenko et al.~(2019).  From their data we find  $H$ =14.9 on UT 2017 September 29 and $H$ = 15.3 on November 11.  These values are systematically fainter than in our data, presumably because these authors used smaller, fixed angle apertures of 10\arcsec~radius.

The spatial distribution of dust in the WIYN and DECam images offers clues about the properties of the ejected dust.  We used the WIYN and DECam images to constrain the particle properties in S5 using the three dimensional dust dynamics model described by Ishiguro et al.~(2007) and used in Hsieh et al.~(2009).  While this model is observationally under-constrained and therefore non-unique, it serves to provide a consistent and physically informative description of the dust properties in a comet.   We assumed that the ejected particles follow a differential power law size distribution, with index $q$ = -3.5.  The ejection speed was assumed to obey $V = V_0 \beta^{u1} r_H^{-u2}$, with $V_0$ = 30 m s$^{-1}$, $u1$ = 0.5, $u2$ = 0.5 and $r_H$ expressed in AU.  The choice of parameters was guided, in part, by prior application of this model to active asteroids (e.g.~Hsieh et al.~2009) and by particle speed measurements at 67P/Churyumov-Gerasimenko (Della-Corte et al.~2016).  Here, $\beta$ is the ratio of the acceleration due to radiation pressure to the local acceleration due to the gravity of the Sun.  It is inversely related to the product $\rho a$ (where $\rho$ is the dust grain density) and numerically, for dielectric spheres, is approximately equal to the inverse of the particle radius expressed in microns, $\beta \sim a_{\mu m}^{-1}$.  The $V \propto \beta^{1/2}$ functional form is applicable to particle ejection by gas drag in the absence of cohesion.  We also assumed that dust particles are ejected symmetrically with respect to the Sun-comet axis in a cone-shape distribution with half-opening angle $w$ = 30\degr.

\begin{figure}
\epsscale{0.99}
\plotone{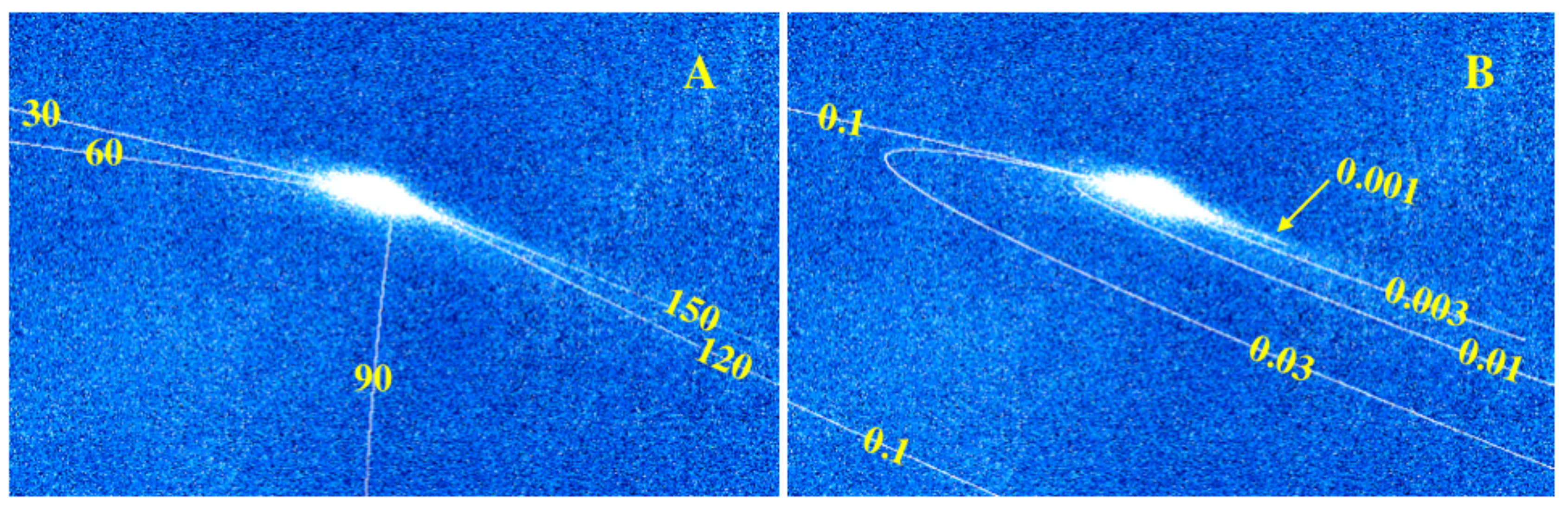}
\caption{A) synchrones computed for ejection 30, 60, 90, 120 and 150 days prior to the date of observation and B) syndynes, showing the paths of particles with $\beta$ = 0.1, 0.03, 0.01, 0.003 and 0.001. \label{synsyn}}
\end{figure}

\begin{figure}
\epsscale{0.99}
\plotone{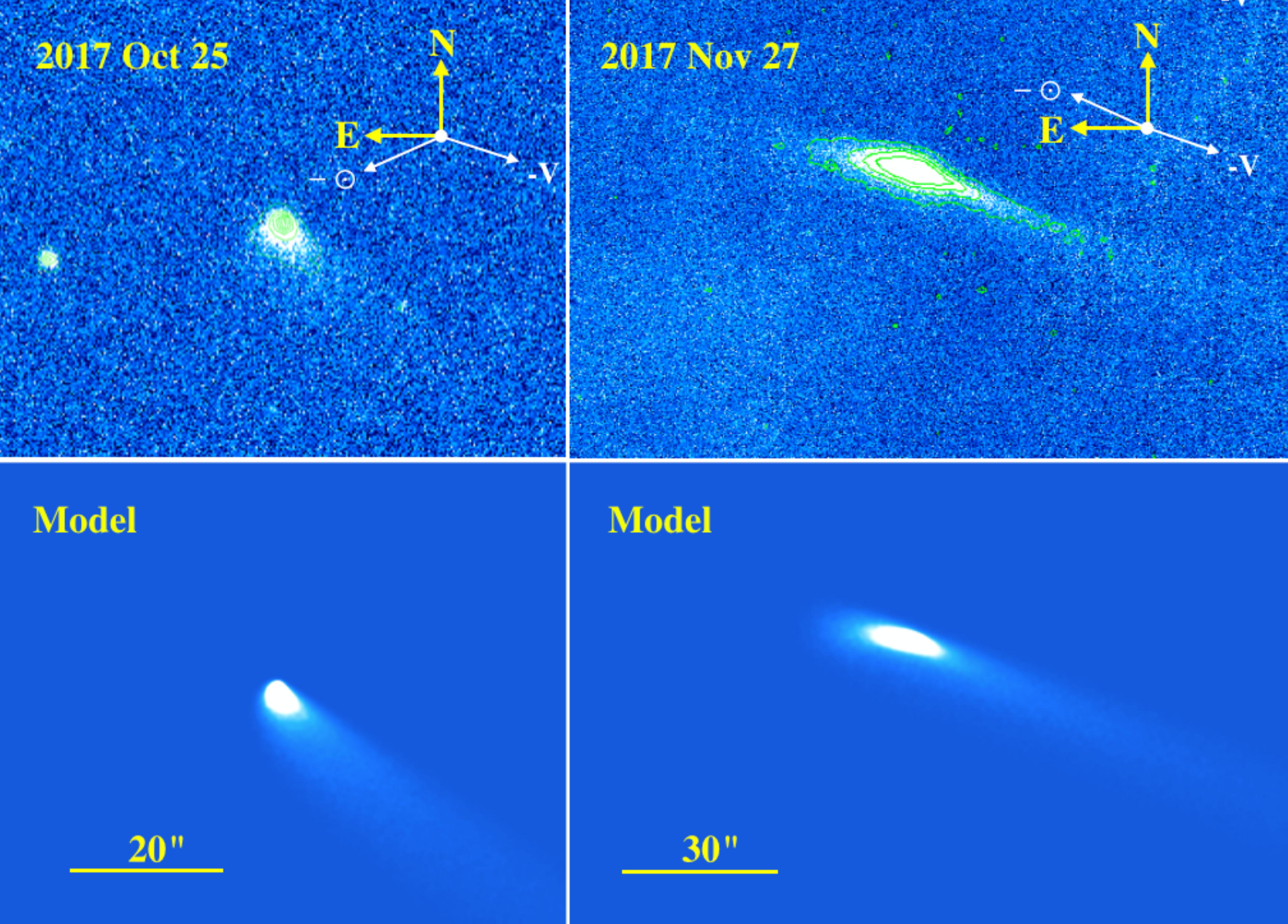}
\caption{Model simulations compared with images from (left) DECam and (right) WIYN.  Adopted model parameters are described in the text.  \label{simulations}}
\end{figure}

With these model parameters, we find from the WIYN and DECam images plausible image-plane solutions for $10^{-4} \le \beta \le 10^{-2}$ (although solutions with $10^{-3} \le \beta \le 10^{-2}$ are almost as good) and steady emission of duration $\Delta t$ = 150 days starting DOY $\sim$ 180 (Figures \ref{synsyn} and \ref{simulations}). The latter corresponds to UT 2017 Jun 30 or about a month prior to the July 27 perihelion, when S5 was at $r_H$ = 2.186 AU.  The solutions for $\beta$ correspond to approximate minimum and maximum particle radii $a_0 = 100~\mu$m and $a_1 = 10^4~\mu$m, respectively.  The model ejection velocities of the particles range from $V \sim 0.2$ m s$^{-1}$ for the smallest $\beta$ (largest $a$) to $V \sim 2$ m s$^{-1}$ for the largest $\beta$ (smallest $a$).  Both of these velocities  are close to the estimated $V_e$ = 0.3 m s$^{-1}$ escape speed from the nucleus but are very slow compared to the $\sim$400 m s$^{-1}$ sound speed in gas sublimating from  the subsolar point on the nucleus.   Low ejection velocities are typical of dust in the active asteroids, partly as a result of their small nucleus sizes, partly of geometrical effects resulting from the small source size on the nucleus (Jewitt et al.~2014) and also indicating that the gas flux is very weak.  

The best-fit parameters from the dust model, when combined with the photometry, yield an estimate of the mass production rate.  Table (\ref{photometry2}) shows that the cross-section increased by $\Delta C_e \sim$ 16$\pm$9 km$^2$ between UT 2017 October 25 and November 27, an interval of $\tau = 3\times 10^6$ s.  In a collection of spheres, the cross-section and the mass are related by $\Delta M_d \sim  \rho \overline{a} \Delta C_e$, where $\overline{a}$ is the mean dust grain radius weighted by the size distribution.  We take $\rho$ = 1000 kg m$^{-3}$ and set $\overline{a} = (a_0 a_1)^{1/2}$ giving  $\overline{a} = 10^{-3}$ m.  Then, the  implied average dust mass loss rate from the nucleus in this interval was $dM_d/dt \sim$ 5$\pm$3 kg s$^{-1}$.  This is strictly a measure of the difference between the rates of production of dust and loss through the outer edge of the photometry aperture, and so constitutes a practical lower limit to the true production rate from the nucleus.  Nevertheless, the derived rate is very comparable to the mass loss rates inferred in other active asteroids using dust photometry (Jewitt et al.~2015).  Most interestingly, the low-speed ejection of large particles, the few kg s$^{-1}$ mass-loss rate and the $\sim$10$^2$ day duration of activity are similar to these quantities measured in the dynamically related, sub-kilometer split active asteroid P/2016 J1-A, J1-B (Moreno et al.~2017, Hui et al.~2017).

If activity had stopped soon after the WIYN observation on UT 2017 November 27 (true anomaly $\nu$ = 41\degr), particles with $\beta  \gtrsim 3\times 10^{-4}$ (i.e.~$a \lesssim$ 3 mm), would be accelerated beyond the field of view of the HST data, explaining their absence.

\begin{figure}
\epsscale{0.95}
\plotone{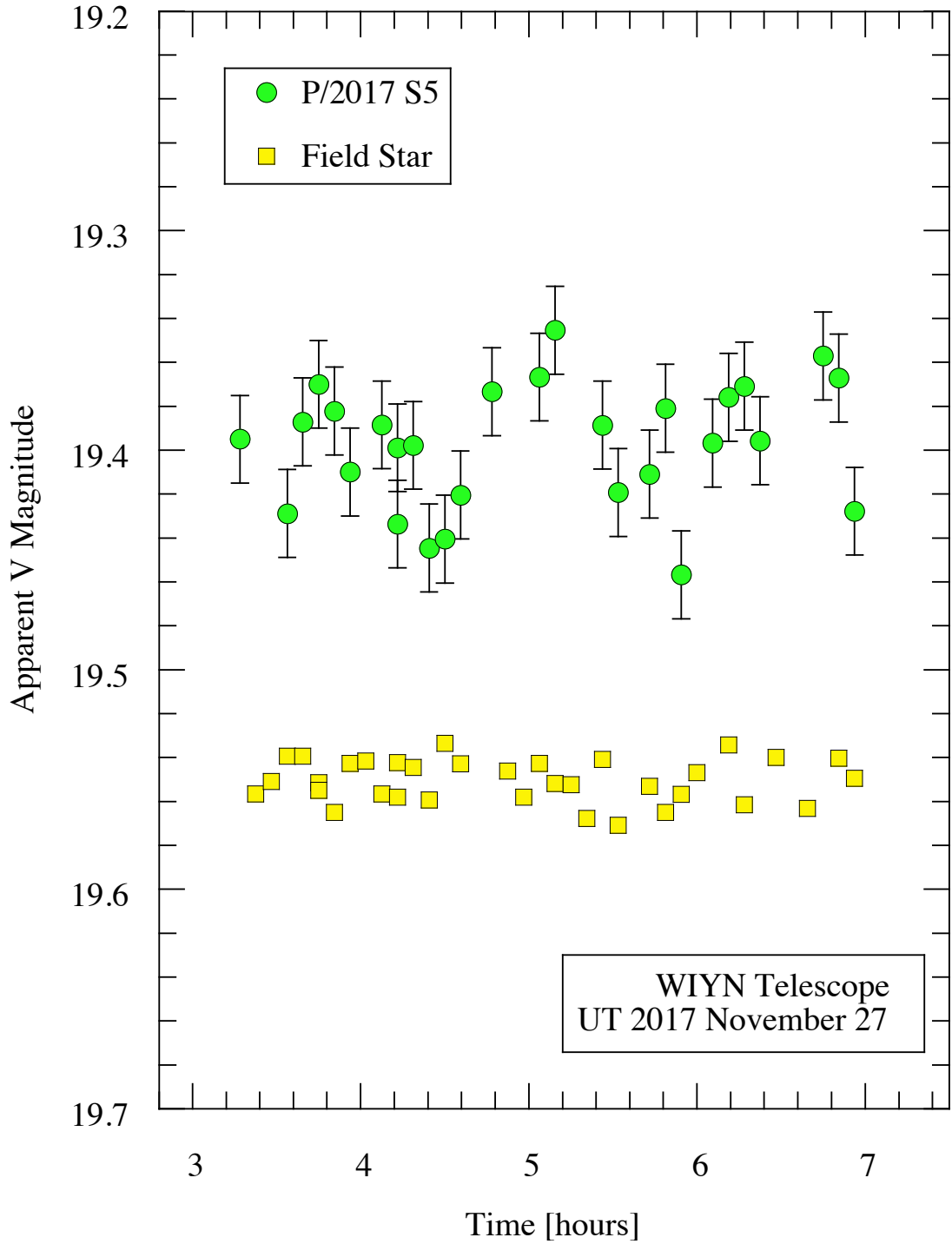}
\caption{Photometry from WIYN (3\arcsec~radius aperture) compared with a field star of comparable brightness.  
\label{wiyn}}
\end{figure}

We also searched the WIYN data for evidence of time-dependent photometric variations.  We find that, over the $\sim$3.6 hour observing interval, S5 does vary by an amount larger than field stars of comparable brightness (Figure \ref{wiyn}).  For example, the standard deviation on the mean of the S5 measurements in the Figure is $1\sigma = $8.6 milli-magnitudes (mmag) from 31 images, while that of the field star (and other stars nearby  of similar brightness) is $1\sigma = $1.7 mmag (34 images).   We considered the possibility that the variation in S5 might be caused by its motion across unseen, fixed  background objects.  However, with instantaneous non-sidereal angular rates of 23\arcsec~hour$^{-1}$ in the WIYN data, the time for a source to cross the full 6\arcsec~diameter of the photometry aperture is only $\sim$ 1/4 hr, whereas the structures in Figure (\ref{wiyn}) have a timescale $\sim$1 hour.  Therefore, we are inclined to believe that the variations are real, although of small range (peak-to-valley $\sim$0.07 magnitudes).  

A phase dispersion minimization estimate of the period in the S5 data gives a best-fit peak-to-peak period $P_L$ = 1.44 hours.  Given the absolute magnitude of the nucleus determined from HST data at large $r_H$ (Table \ref{photometry2}), the nucleus contribution to the  WIYN magnitudes in Figure (\ref{wiyn}) is only $\lesssim$5\%.  Therefore,  the variations in the WIYN photometry cannot be uniquely reflective of the shape of the underlying nucleus (which, if it were the case, would indicate a rotational period $2P_L$ = 2.88 hour, because of rotational symmetry).  Instead, it is more likely that the variations are produced  indirectly, perhaps by periodic illumination of an active region on a nucleus rotating at period $P_L$.  Whether the period is $P$ or $2P$, however, the WIYN data raise the possibility that S5 is rotating close to rotational instability.   For example,  rotation with a (2 hour) period similar to that of S5 is implicated in the fragmentation of the $<$275 m radius nucleus of 332P/Ikeya-Murakami (Jewitt et al.~2016).   Pending the acquisition of better photometry needed to confirm that the variations in Figure (\ref{wiyn}) are periodic, however, we leave open the role of  rotation in affecting the activity of S5. 

\subsection{Mechanisms}
The distribution of the dust bears no simple relation to any of the synchrones plotted in  Figure (\ref{synsyn}), and so is inconsistent with impulsive ejection. For example, while the 120 and 150 day isochrones approximately match the position angle of the tail to the west of the nucleus, they leave the (evidently much younger) dust to the east unexplained.  Conversely, the dust to the east is well-matched by the 30 and 60 day isochrones, but these do not fit the long tail to the west.  Evidently, emission over a long period is required to account for the observed spatial distribution of the dust.   On this basis, we discount the possibility that activity in S5 could be caused by a sudden ejection of dust of the type expected from an impact.    

On the other hand, long-duration dust emission can be most naturally explained by sublimation, which we identify as the leading candidate  for the dust emission mechanism in S5.  
We solved the energy balance equation (neglecting conduction) to find that, at $r_H$ = 2.3 AU, the mass flux from an exposed water ice surface sublimating in equilibrium with sunlight varies from $f_s = 3\times 10^{-6}$ kg m$^{-2}$ s$^{-1}$ at the isothermal temperature ($T$ = 175 K) to $f_s = 6\times 10^{-5}$ kg m$^{-2}$ at the subsolar point ($T$ = 192 K).   Therefore, a  production rate of $\sim$5 kg s$^{-1}$ could be supplied by sublimation from exposed ice  covering only $\sim$0.1 km$^2$ near the maximum possible temperatures at the subsolar point, rising to 1.5 km$^2$ if the ice is globally distributed and isothermal.   For reference, the total surface of a 0.45 km radius sphere is 2.5 km$^2$, so that the active fraction for sub-solar sublimation is $f_A \sim$ 0.04, rising to $f_A \sim$ 0.6 in the isothermal case.  The former would be typical of the nuclei of short-period comets while the latter would be unusually (but not uniquely) large (A'Hearn et al.~1995).  We also note that the osculating eccentricity of S5 is currently near a maximum, and the perihelion near a minimum, so raising the prospect of enhanced sublimation, following the alignment effect noted by Kim et al.~(2018).

At the above specific sublimation rates, the time needed for ice to sublimate over a distance comparable to the nucleus radius, $r_e$, is $\tau = r_e \rho /(f_s f_t)$, where $\rho$ = 1000 kg m$^{-3}$ is the assumed density and $f_t$ is the fraction of each orbit spent in the active state. For example if, as our data suggest, S5 is active for $\sim$0.5 year in each $\sim$5 year orbit, we have $f_t \sim$ 0.1.   Substituting, we find  7.5$\times 10^{10} \le \tau \le 1.5\times10^{12}$ s (roughly a few 10$^3$ to a few 10$^4$ years).  These timescales are very short compared to the 7$\pm$2 Myr dynamical lifetime of the Theobalda family (Novakovic 2018) showing that, as in active asteroids generally, the activity must be transient.  The upper limit to the rate of recession of a sublimating ice surface at the subsolar point is $f_s/\rho \lesssim 10^{-7}$ m s$^{-1}$.  Over the $\Delta t\sim$ 150 days of activity, an ice thickness  $f_s \Delta t/(\rho) \lesssim$ 0.8 m could be lost.  By $r_H$ = 3.156 AU, the distance of S5 in the HST observations, the subsolar sublimation rate would fall to  1.4$\times 10^{-5}$ kg m$^{-2}$ s$^{-1}$, only 5$\times$ smaller than at perihelion and still capable of generating a detectable coma.  Isothermal sublimation falls faster (by a factor $\sim$100) over the same distance range.  The absence of coma  at $r_H$ = 3.156 AU could indicate that the ice is sublimating through a thin, refractory surface crust, so that its temperature and sublimation flux are lower than would be the case if the ice were exposed at the surface.  This would also naturally account for a steep $r_H$ dependence of the sublimation mass flux.  Shadowing on the nucleus, caused by surface topography and the changing illumination geometry (e.g.~the true anomaly changed by $\sim$60\degr, c.f.~Table \ref{geometry}), could also be responsible.   

The conduction timescale for a body of radius $r_e$ is $\tau_c = r_e^2/\kappa$, where $\kappa$ (m$^2$ s$^{-1}$) is the thermal diffusivity of the material.  Solid rocks have $\kappa \sim 10^{-6}$ m$^2$ s$^{-1}$ while porosity decreases $\kappa$.  With $\kappa = 10^{-6}$ m$^2$ s$^{-1}$ the conduction timescale for a 450 m nucleus is $\tau_c \sim$ 7000 years.  In order for $\tau_c$ to exceed the dynamical lifetime of the Theobalda family would require $\kappa < r_e^2/\tau_d$, or $\kappa < 4\times10^{-10}$ m$^2$ s$^{-1}$, which is small even compared to the thermal diffusivity of the Lunar regolith. Therefore, it is reasonable to expect that even the core of S5 has equilibrated to a temperature determined by its orbit, whatever its initial value.  We estimate core temperature $T \sim$ 156 K, at which water ice could remain stable for the dynamical age of the Theobalda family.  

Two other active asteroids, the recently-split pair P/2016 J1-A, J1-B, may also be  members of the Theobalda family (Hsieh et al.~2018, Novakovic 2018).  Physically, the two components of P/2016 J1 resemble S5 in that they are small (sub-kilometer) bodies, which ejected large dust particles ($a \gtrsim 100~\mu$m) for hundreds of days, starting before perihelion. Production rates,  $\lesssim$ 1 kg s$^{-1}$,  and  ejection speeds, $\sim$0.5 m s$^{-1}$, were also comparable (Moreno et al.~2017, Hui et al.~2017).  A plausible interpretation is that S5 is another fragment of the Theobalda parent body that has been able to retain sub-surface ice since the disruption event and which was recently triggered either by a small impact, a landslide or another surface instability.  This interpretation is eminently  testable; if ice is responsible then all three objects should reactivate at their next perihelia, expected 2022.2 for P/2016 J1-A and J1-B, and in 2023.3 for S5.   

As noted above, oscillations in the brightness of S5 determined from the WIYN data (Figure \ref{wiyn}) suggest rapid nucleus rotation. Future observations should target the potential role of rotation in the three bodies, all of which  are small enough that mass-loss torques are  easily capable of accelerating the  spin. For example, Equation (3) of Jewitt et al.~2016 gives an e-folding spin-up timescale of only  $\tau_s \lesssim$ 10$^3$ years for $r_n$ = 450 m, mass loss rate $dM/dt =$ 5 kg s$^{-1}$, moment arm $k_T$ = 0.005 and period $P$ = 1.4 hours.  Regardless of the uncertainties in the adopted parameters, $\tau_s$ is very small compared to the Theobalda family age, $\tau_d = 7\pm2$ Myr.    Indeed, the survival of S5 against rotational break-up for a time $\tau_d$ requires that the body be largely inactive, with a duty (``on/off'' ratio) cycle $\tau_s/\tau_d \lesssim 10^{-4}$.   Other members of the Theobalda family should also be examined for evidence of rapid rotation and  sublimation-driven mass loss.

\clearpage

\section{SUMMARY}

\begin{enumerate}

\item The nucleus of P/2017 S5 has absolute magnitude $H$ = 18.91$\pm$0.03.  With assumed geometric albedo $p_V = 0.06\pm0.02$, the equivalent circular radius is $r_e = 450_{-60}^{+100}$ m. 

\item Impact and other impulsive  origins are ruled out for S5 by the spatial distribution of the dust, which requires a persistent source (lasting 150 days, or more).  The scattering cross-section is concentrated in large particles ($10^{-4} \le \beta \le 10^{-2}$, radii $10^2 \le a \le 10^4~\mu$m), with a mass production rate $\sim$5$\pm$3 kg s$^{-1}$.  In addition, the dust particles are ejected very slowly, with speeds comparable to the $V_e$ = 0.3 m s$^{-1}$ gravitational escape speed from the nucleus.

\item Equilibrium sublimation of recently exposed water ice from as little as 0.1 km$^2$ ($\sim$4\% of the nucleus surface) can account for the duration and the magnitude of the mass loss.  

\item  Time-series photometry provides tentative evidence  for rapid rotation of the nucleus (lightcurve period $\sim$1.4 hour), consistent with the small nucleus size and resulting rapid spin-up timescale under the action of sublimation torques.  However,  the role played by rotation in the observed activity remains unclear.

\item The available data are consistent with P/2017 S5 being an ice-containing main-belt comet, sharing both physical and dynamical similarities with the split active asteroid pair P/2016 J1-A and J1-B.

\end{enumerate}

\acknowledgments
We thank Man-To Hui and the anonymous reviewer for comments on the manuscript.  Based on observations made under GO 15342  with the NASA/ESA Hubble Space Telescope, obtained at the Space Telescope Science Institute,  operated by the Association of Universities for Research in Astronomy, Inc., under NASA contract NAS 5-26555.   YK acknowledges funding from Grant 50 OR 1703 by the German Aerospace Center (DLR).



{\it Facilities:}  \facility{HST, WIYN}.

\clearpage


\clearpage

\begin{deluxetable}{lccrrrccccr}
\tablecaption{Observing Geometry 
\label{geometry}}
\tablewidth{0pt}
\tablehead{\colhead{UT Date} & \colhead{Tel\tablenotemark{a}} & \colhead{DOY\tablenotemark{b}} & \colhead{$\nu$\tablenotemark{c}} & \colhead{$r_H$\tablenotemark{d}} & \colhead{$\Delta$\tablenotemark{e}}  & \colhead{$\alpha$\tablenotemark{f}} & \colhead{$\theta_{- \odot}$\tablenotemark{g}} & \colhead{$\theta_{-V}$\tablenotemark{h}} & \colhead{$\delta_{\oplus}$\tablenotemark{i}}   }

\startdata

2017 Oct  25  & CTIO & 298 & 30.6 & 2.253 & 1.297 &   9.2 & 112.3 & 252.7 &  -5.9 \\
2017 Nov 27  & WIYN & 331 & 41.2 & 2.315 & 1.563 & 19.3 &  75.5 &  252.6 &  -1.0 \\
2018 Sep 08 
& HST & 616 & 107.5  & 3.156 & 3.672 & 14.6 & 276.1 & 290.6 &  -3.2 \\

\enddata

\tablenotetext{a}{Telescope: CTIO = 4.0 m Cerro Tololo Inter-American Observatory, WIYN = 3.5 m Wisconsin Indiana Yale NOAO, HST = 2.4 m Hubble Space Telescope}
\tablenotetext{b}{Day of Year, DOY = 1 on UT 2017 January 01}
\tablenotetext{c}{True anomaly (degree)}
\tablenotetext{d}{Heliocentric distance, in AU}
\tablenotetext{e}{Geocentric distance, in AU}
\tablenotetext{f}{Phase angle, in degrees}
\tablenotetext{g}{Position angle of projected anti-solar direction, in degrees}
\tablenotetext{h}{Position angle of negative projected orbit vector, in degrees}
\tablenotetext{i}{Angle from orbital plane, in degrees}

\end{deluxetable}

\clearpage

\begin{deluxetable}{lcccccc}

\tablecaption{HST Fixed-Aperture Photometry\tablenotemark{a} 
\label{photometry1}}
\tablewidth{0pt}

\tablehead{ \colhead{UT Start\tablenotemark{b}} & Decimal UT & $V$ & $H$  }
\startdata

20:20:53	& 20.348	& 	24.99$\pm$0.07	& 	18.85$\pm$0.07\\
20:29:01	& 20.484 	& 	25.02$\pm$0.07	& 	18.88$\pm$0.07\\
20:37:09	& 20.619 	& 	25.12$\pm$0.07	& 	18.98$\pm$0.07\\
20:45:17	& 20.755	& 	25.08$\pm$0.07	& 	18.94$\pm$0.07\\
20:53:25	& 20.890	& 	25.05$\pm$0.07	& 	18.91$\pm$0.07\\

22:04:20	& 22.072	& 	25.21$\pm$0.07	& 	19.07$\pm$0.07\\
22:12:28	& 22.208	&       24.97$\pm$0.07	& 	18.83$\pm$0.07\\
22:20:36	& 22.343	& 	25.07$\pm$0.07	& 	18.93$\pm$0.07\\
22:28:44	& 22.479	& 	24.98$\pm$0.07	& 	18.84$\pm$0.07\\

\hline
MEAN    	&		&	25.05$\pm$0.03	& 	18.91$\pm$0.03	\\

\enddata

\tablenotetext{a}{Aperture radius 0.2\arcsec.}
\tablenotetext{b}{UT 2018 September 08}
\end{deluxetable}

\clearpage

\begin{deluxetable}{lcccl}

\tablecaption{Averaged Photometry 
\label{photometry2}}
\tablewidth{0pt}

\tablehead{ \colhead{UT Date} & Aper\tablenotemark{a} & V  & $H$ & $C_e$ [km$^2$]\tablenotemark{b} }
\startdata

2017 Oct 25	& Fixed & 17.63$\pm$0.05	& 	14.69$\pm$0.10	&   31$\pm$3  \\
2017 Nov 27	& Fixed & 18.02$\pm$0.10	& 	14.25$\pm$0.20	&   47$\pm$9 \\
2018 Sep 08	& Point & 25.05$\pm$0.03	& 	18.91$\pm$0.03	&   0.65$\pm$0.02  \\

\enddata

\tablenotetext{a}{Aperture used; Fixed = radius 25,000 km at the distance of S5, Point = S5 unresolved, aperture radius 0.2\arcsec~= 530 km}
\tablenotetext{b}{Cross-section computed from Equation (\ref{invsq}) assuming geometric albedo $p_V$ = 0.06}

\end{deluxetable}

\clearpage

\clearpage

\clearpage

\clearpage

\clearpage

\clearpage

\end{document}